\documentstyle[12pt,psfig]{article}
\begin{document}

\vskip 1truecm
\rightline{Preprint  PUPT-1698}
\rightline{ e-Print Archive: hep-ph/9705248}
\vspace{0.2in}
\centerline{\Large Computing the Strong Sphaleron Rate} 
\vspace{0.3in}
\centerline{\Large Guy D. Moore\footnote{e-mail:
guymoore@puhep1.princeton.edu }
}
\medskip

\centerline{\it Princeton University}
\centerline{\it Joseph Henry Laboratories, PO Box 708}
\centerline{\it Princeton, NJ 08544, USA}

\vspace{0.2in}

\centerline{\bf Abstract}
We measure the diffusion constant for Chern-Simons number for classical,
lattice SU(3) Yang-Mills theory, using a generalization of the
topological definition of
Chern-Simons number developed recently by Moore and Turok.  The diffusion
constant is much larger than that for SU(2), even before the ratio of
coupling constants has been accounted for, which implies that 
chiral quark number is efficiently destroyed by strong processes during
the electroweak phase transition.  For the physical value of
$\alpha_s$ we estimate the decay time for
chiral quark number to be about $80 / T$, although various systematics
make this number uncertain by about a factor of 2.

\begin{verse} 
\end{verse}

\section{Introduction}
\label{introduction}

Baryon number is violated in the standard model and
there has been a growing interest in trying 
to understand mechanisms which might use
this violation to generate the baryon asymmetry
of the universe during the cosmological electroweak
phase transition.

All the needed ingredients for generating a
baryon asymmetry \cite{Sakharov} are present; the
violation of baryon number shuts off abruptly while
the plasma is out of equilibrium
due to the motion of a bubble wall (phase 
boundary), and if there is $C$ and $CP$ violation
then these conditions can give rise to
a net baryon number generation.
Interest in this scenario has focused
especially on a particularly efficient
mechanism in which the $CP$ violation, in
the form of spatially varying Higgs condensate
phases on the bubble wall surface,
generates a chiral top quark asymmetry \cite{various}.
This can be transported by particle diffusion \cite{two,CKNdiffusion}
into the symmetric electroweak phase; the left
handed quark number then biases SU(2) winding number
changing transitions (``sphalerons'') which
generates a net baryon number.

A complication to this scenario is that chiral
quark number is damped by SU(3) color winding number
changing events \cite{SSearly,Giudice}, similar to the
phenomenon responsible for the spontaneous
breaking of chiral symmetry in QCD.  To be more
concrete, if there is a chemical potential $\mu$
for chiral quark number, then the total chiral quark
number density (left handed quarks minus their antiparticles
minus righthanded quarks, plus their antiparticles)
is 
\begin{equation}
Q_5 \simeq \sum_{\rm species} \frac{\mu T^2}{12} 
	= 8 N_c N_F \frac{\mu T^2}{12} \, ,
\end{equation}
where the $8$ comes from summing over up and down types, particles
and antiparticles, and the two chiralities, and the $T^2/12$ is
the (leading order in $\mu$ and $g^2$) change in number density due to
the chemical potential for an ultrarelativistic fermionic
species.  ($N_c=3$ is the number
of colors and $N_F=3$ is the number of fermion generations.)  The 
free energy liberated by a strong sphaleron event is $\mu$ for
each left handed particle destroyed or righthanded particle 
created, which equals $4 N_F$, since in each generation there is
a creation of a righthanded particle and a destruction of a lefthanded
particle, both for up and down type quarks.  The rate per unit volume
of strong sphaleron transitions will then be
$4 N_F (\mu/T) \Gamma_{ss}$, where $\Gamma_{ss}$, the linear
response coefficient of strong sphalerons to a chemical
potential, equals half the diffusion constant per unit volume
of SU(3) Chern-Simons number, by a fluctuation
dissipation relation \cite{KhlebShap,RubakShap2,Moore1}.  Since each
transition changes chiral quark number by $-4 N_F$, one finds\footnote{
The derivation here follows that in \protect{\cite{Giudice}} except
that they miss the factor of $N_c$ in the relation between $Q_5$ and
$\mu$, so their rate constant for the decay of $Q_5$ is 3 times too large.}
\begin{equation}
\frac{dQ_5}{dt} = (4 N_F)^2 \frac{\mu}{T} \Gamma_{ss} 
	= -\frac{24 N_F}{N_c} Q_5
	\frac{\Gamma_{ss}}{T^3} \, .
\end{equation}
The time constant for the decay of $Q_5$ is therefore
\begin{equation}
\tau = \frac{T^3}{24 \Gamma_{ss}} \, .
\end{equation}

Even without calculation we know that $\Gamma_{ss}$ is much larger than
the corresponding weak sphaleron rate $\Gamma_{ws}$, 
because SU(3) contains SU(2) and
the strong coupling is larger (so nonperturbative physics sets in on
a shorter length scale).  Hence chiral quark number in the symmetric
electroweak phase decays mainly through strong phenomena.  If the time
constant is shorter than the typical time a quark reflected from or
moving off of the bubble wall spends in the symmetric phase before the
wall catches up with it, then the strong sphaleron rate will be relevant
and will reduce the baryon number generation.  In this case we may
only need to know the ratio of the strong and weak sphaleron rates to
determine the baryon number generated.  If $\tau$ is 
shorter than the time it takes a particle to get from the middle of the
bubble wall to the symmetric phase, then the chiral quark number may be
destroyed before it reaches an environment where it can generate baryon
number, and the strong sphalerons will qualitatively 
reduce the production rate for baryon number.  (It is 
important to treat all relevant processes, such as the difference in
diffusion constants between right and left handed quarks, 
and the conversion between chiral top quark number and Higgs particle 
abundance (net Higgs particle hypercharge) due to the top quark 
Yukawa coupling \cite{CKNdiffusion}; these
might change this picture somewhat.)
In any case it seems that the investigation of $\Gamma_{ss}$ is well
motivated.  

It is conventional to write $\Gamma$ in
terms of a dimensionless constant $\kappa$,
\begin{equation}
\Gamma_{ss} = \kappa_{ss} \alpha_s^4 T^4 / 2 \, ,
\label{naivescaling}
\end{equation}
where the factor of $2$ is because in conventional usage 
$\kappa_{ss}$ denotes the diffusion constant for $N_{CS}$, 
not the response coefficient.  $\kappa_{ss}$ may depend
nontrivially on couplings and the particle content.  One should take
$\alpha_s$ at a renormalization point on order the temperature 
scale.  Beyond leading
order, $\kappa$ must have a logarithmic renormalization point
dependence; we will discuss the ``best'' renormalization point in the
next section.

A first attempt to compute $\Gamma_{ss}$ was
made in \cite{Moore1}, using the classical approximation for the infrared
dynamics first suggested in \cite{GrigRub} and a numerical implementation
of Yang-Mills theory and of Chern-Simons number developed and used
in \cite{Ambjornetal,AmbKras}.  Unfortunately, this definition of
Chern-Simons number suffers from lattice artifacts, leading to a
spurious ultraviolet signal and an incorrect normalization of the response
to the real, infrared winding number change.  The lattice implementation
of Yang-Mills theory also requires perturbative corrections to the
tree level match between
lattice and physical length scales \cite{Oapaper}.  
It is known in the case of SU(2) that both corrections are numerically
important.  The SU(2) diffusion constant without these corrections
gives a lattice spacing independent $\kappa$ \cite{AmbKras}, 
which is qualitatively
different from what is expected on theoretical grounds.  Arnold, Son,
and Yaffe have argued that the interaction of the infrared
modes responsible for winding number change with ``hard'' short wavelength
excitations is essential, and because of it 
$\kappa$ should depend on the product
of the plasma frequency $\omega_{pl}$ and the inverse nonperturbative
length scale $l_{np} \propto 1/g^2 T$ as
\begin{equation}
\kappa \propto ( \omega_{pl} l_{np} )^{-2} \, ,
\end{equation}
at least when $( \omega_{pl} l_{np} )^{2} \gg 1$ \cite{ArnoldYaffe}.  
Since $\omega_{pl}$ on the lattice depends on the lattice spacing 
$a$ as $a^{-1/2}$ \cite{Arnoldlatt}, this
implies that $\kappa$ should be proportional to $a$; and for the physical
quantum system $\omega_{pl} \sim gT$, so $\kappa_{ws} \propto \alpha_w$.

Recently a topological technique for measuring winding number change
on the lattice has been developed for classical SU(2) gauge theory 
\cite{slavepaper}.  Using it, and applying the corrected matching between
lattice and physical length scales, reveals that $\Gamma_{ws}$ does 
depend on lattice spacing, verifying that previous results were 
contaminated with lattice artifacts.  The coarsest lattices used there were
insufficient to achieve $( \omega_{pl} l_{np} )^2 \gg 1$, and the
dependence was (therefore?) weaker than linear with $a$; but
we will assume here that the reasoning of Arnold, Son, and Yaffe is 
correct and that for sufficiently large hard thermal loop effects the rate
does scale with $( \omega_{pl} l_{np} )^{-2}$.

Because $\Gamma$ depends on the physics of hard modes, which is definitely
not reproduced correctly on the lattice \cite{Smilga}, 
we do not know how precisely to convert $\Gamma$ measured there 
into $\Gamma$ in the quantum theory.  This problem was addressed
recently by Arnold, who argues that a reasonably accurate conversion
is possible \cite{Arnoldlatt}.  The small $a$ limit of the 
ratio of $\Gamma$ for SU(2) and
SU(3) should also be insensitive to this problem, since the distortion
of the hard thermal loops is common to the two lattice theories.
Hence we can compute the ratio $\Gamma_{ss}/\Gamma_{ws}$ on the
lattice with smaller systematics.  The
purpose of this letter is to extend the results of \cite{Oapaper}
and the technique of \cite{slavepaper} to SU(3), and to use them
to compute $\Gamma_{ss}$ on the lattice, and
hence to find $\Gamma_{ss}/\Gamma_{ws}$ and to estimate $\Gamma_{ss}$. 

\section{Diffusion constant:  General discussion}
\label{defsphal}

Most analytic work on $\Gamma$ has discussed the broken phase of
Yang-Mills Higgs (YMH) theory.  In this case the gauge connection is
usually close to some topological vacuum, and the diffusion
rate is controlled
by the free energy of configurations midway between vacua,
``sphalerons.''  This way of viewing things can also be useful in the
symmetric phase or in pure Yang-Mills (YM) theory, where the 
diffusion rate will
also depend on how often the system is straddling between vacua.

Consider classical YM or YMH theory in a fixed finite volume and
regulated at a length scale much smaller than $1/g^2 T$.  (We are
thinking of lattice regulation, but we will use continuum notation
here for convenience.)  We can
define the closest vacuum to the spatial connection $\vec{A}$ at time
$t$ by ``cooling'' \cite{Hetrick}.  That is, consider evolving 
$\vec{A}$ under (gauge invariant) straight dissipative dynamics,
\begin{equation}
\frac{ \partial {\vec{A}(x,t,\tau)}}{\partial \tau} = - \frac{
	\partial H(A(t,\tau)) }{\partial {\vec{A}(x,t,\tau)}} \, ,
	\quad A(x,t,0) = A(x,t) \, ,
\end{equation}
with $H$ the Hamiltonian.  (The cooling time $\tau$ is not to be confused
with the decay rate for chiral quark number discussed in the
introduction.)  Such cooling was recently used as a
technique to improve the local operator method for tracking $N_{CS}$
\cite{AmbKras2}.  At sufficiently large $\tau$ the connection will
settle into a vacuum configuration, and we define this to be the
nearest vacuum.  

It will sometimes occur that there is a time $t_{\rm
sph}$ such that the nearest vacuum at
time $t_{\rm sph} + \epsilon$ has a different winding number than the
vacuum at time $t_{\rm sph} - \epsilon$.\footnote{This statement also
has a rigorous gauge invariant meaning, even on the lattice, if the
lattice spacing is not too coarse:  the winding number difference of
the vacua is defined as the winding number of the gauge
transformation carrying one to the other,
measured using the algorithm of Woit \cite{Woit}, or, for SU(N), the
technique developed in the next section.}
At time $t_{\rm sph}$ the
dissipative evolution will never get to a vacuum state, and just
before and after, it will take a very long time.  We can define the
system to be ``in a sphaleron'' if the cooling time $\tau$ required to
get close to vacuum exceeds
some threshold $\tau_{\rm thresh}$.  Here ``close to vacuum'' can be
given a rigorous definition, eg the total action of the
remaining cooling path to the vacuum is less than $(\pi/g)^2$.

The probability that the system
is ``in a sphaleron'' depends only on $\tau_{\rm thresh}$ and the
thermodynamics of the spatial connections.  Further, if we define 
\begin{equation}
E^a_i(x,t,\tau) = [ D_0 \, , \, D_i ]^a ( x, t , \tau )
\label{Edef}
\end{equation}
(with $D_0$ the covariant $t$ derivative), then since
the distribution of values for 
$E(t,\tau=0)$ is a thermodynamic property, and since $\partial E /
\partial \tau$ depends only on the connections and on $E$,
\cite{AmbKras2}
\begin{equation}
\frac{\partial E^a_i(x,t,\tau)}{\partial \tau} =  E^b_j(x,t,\tau) 
	\frac{\partial^2 H(A(t,\tau))}{\partial A^a_i(x,t,\tau)
	\partial A^b_j(x,t,\tau)} \, ,
\end{equation}
then $E(t,\tau)$ is also distributed according to thermodynamics.
If we define the sphaleron narrowly
enough that the system typically remains ``in a sphaleron'' for a time
short compared to the inverse plasma frequency, so $E(t,0)$ does not change
much from the beginning to the end of the sphaleron event, then the
length of a sphaleron event will also depend only on $\tau_{\rm
thresh}$ and on thermodynamics, and the total spacetime density of
sphaleron events will be a thermodynamic property, depending on the
thermodynamics of $\vec{A}$ and $\vec{E}$ alone.
We have not shown here that this spacetime density has a good large
volume limit, but we will assume this to be the case.

Since the spacetime density of sphaleron events depends only on
thermodynamics, we know that the quantum theory value is reproduced by
the classical theory in the $a \rightarrow 0$ limit, with corrections
due to the thermodynamics of the $\vec{A}$ fields which are
$O(\alpha^2)$ if we use the dimensional reduction calculation
\cite{KLRS} to
establish the value of the coupling constant of the 3-D theory.  There
may also be $O(\alpha)$ corrections in the thermodynamics of the
${\vec E}$ fields, ie in the relation between time scales
\cite{TangSmit}, which have not been calculated.  Further, since the
only length scale in the thermodynamics of the $\vec{A}$ and $\vec{E}$
fields is $1/(g^2 T)$, the spacetime density of sphaleron events can
be written as $\kappa_1(N_c) \times ( \alpha T )^4$, with $\kappa_1(N_c)$ a
pure number.  In YMH theory, $\kappa_1$ also depends on the Higgs
potential parameters $x$ and $y$.  It should approach the YM theory
value in the limit of large positive $y$, deep in the symmetric
phase, and it becomes exponentially small at large negative $y$, deep
in the broken phase.

Each sphaleron event changes the winding number of the nearest vacuum
by $\pm 1$.  If the signs of each change were independent, then
$\Gamma$ would equal $\kappa_1(N_c) \alpha^4 T^4 / 2$.  But the
signs will in general be correlated, $\Gamma = \kappa_2 \kappa_1(N_c) 
\alpha^4 T^4 / 2$, where $\kappa_2$ describes the degree of
correlation in the signs of sphaleron events, and depends on the
dynamics.  In particular, Arnold, Son, and Yaffe argue that plasma
oscillations will make the system
go back and forth through sphalerons of opposite sign.  On short time
scales the motion of infrared magnetic fields will be oscillatory, and
on long time scales it will be overdamped.  They conclude that the
system will go through on order
$(\omega_{\rm pl}/g^2T)^2 \sim 1/(g^2 \hbar)$ sphaleron events 
per permanent winding number change, so $\kappa_2 \sim g^2 \hbar$
\cite{ArnoldYaffe}.  In the classical
theory, the role of $\hbar$ is played by the regulator scale, and
$\kappa_2 \propto g^2 a T/4 \equiv \beta_L^{-1}$, at least for large
$\beta_L$ \cite{Arnoldlatt}.

To determine $\Gamma$ correctly, it is necessary to count winding
number changes correctly; to get the
thermodynamics, and hence $\kappa_1$, right; and to get the dynamics,
and hence
$\kappa_2$, right.  The first two problems are separate from the
third, and we deal with them in the next section.  Getting the
dynamics right is harder, and in our opinion this problem has not been
solved.  However, the work of Arnold \cite{Arnoldlatt} suggests that
lattice results can be converted to continuum results with fairly
modest systematic error, and we will use his matching here.

\section{From SU(2) to SU($N_c$)}
\label{section3}

There are no complications in extending 
the standard Kogut-Susskind implementation of 
3+1 dimensional SU(2) Yang Mills theory \cite{KogutSusskind} to SU(3),
and the thermalization algorithm for the SU(3) case was developed in
\cite{Moore1}.  What remains is to extend the one loop matching of the
thermodynamics of lattice and continuum systems, and to extend the 
topological tracking of winding number, from SU(2) to SU(3) (or
SU($N_c$)).

\subsection{thermodynamics}

We deal first with the thermodynamics.  
As we discussed in the last section, it is only the thermodynamics of
the spatial connections (and of the $E$ fields) which are important;
so our goal is to make sure that
the thermodynamics of spatial connections at finite lattice
spacing $a$ are as close as possible to the continuum thermodynamics.
Here and throughout we will use the notation of \cite{AmbKras,Oapaper}.
We will not attempt to make this section self-contained; the
reader is referred to \cite{Oapaper} for details on the approach.  All
we do here is generalize to SU($N_c$) the 
gauge field part of the SU(2) calculation done there.
The details of this section are not important in what follows, only 
the final result, so the uninterested reader can skip to the next subsection.

The thermodynamics of the real time system we are considering
are determined by the path integral
\begin{eqnarray}
Z & = & \int {\cal D}A_i {\cal}A_0 \exp(- \beta_L H_L ) \, , \\
H_L & = & \sum_{x, i<j} \left( \frac{N_c}{2} 
	- \frac{1}{2} {\rm Re \; Tr} U_{ij}(x) \right) 
	+ \nonumber \\
	& & + \sum_{x,i} \frac{1}{2} (U_i(x) A_0(x+i) U^{\dagger}_i(x) 
	- A_0(x))^2 + \sum_x \frac{m_D^2}{2} A_0^2(x) \, ,
\label{Hamiltonian}
\end{eqnarray}
with the bare Debye mass $m_D^2 = 0$ \cite{AmbKras}.
Here $U_{ij}(x)$ is the elementary plaquette which extends from $x$ in
the $i,j$ directions.  The gauge coupling has been absorbed into $\beta_L$
which (at tree level) equals\footnote{For $N_c \neq 2$ this notation 
differs from that usually used in 4 dimensional lattice QCD, 
which uses $1 - (1/N){\rm Re \; Tr}$.} $\beta_L = 4/(g^2 a T)$.
We want to improve the Hamiltonian so the thermodynamics produced by this
partition function match more accurately those of the continuum system.
The idea is that, since the lattice and continuum theories
only differ strongly in the ultraviolet, one should
compute the influence of ultraviolet modes on the infrared physics 
perturbatively, in the lattice and continuum theories, and find the 
difference (which is free of infrared divergences).  
Because the infrared length scale is well separated from the length
scale where the lattice and continuum theories significantly differ,
the difference can be written as an operator product expansion, and only
the super-renormalizable terms are needed.  One compensates
for these terms by making shifts in the wave functions and
couplings of the theory, thereby correcting the lattice theory in the
infrared for its ultraviolet differences from the continuum theory.  
Because the theory is super-renormalizable, one loop perturbative
corrections are $O(a)$,
leaving the infrared behavior of the lattice and continuum theories 
matching up to $O(a^2)$.

The exception to this rule is dimension 2 operators, where the one loop
contribution is $O(1/a)$ in physical units, which in lattice units is
$m^2 \sim \beta_L^{-1}$, and
a full $O(a)$ correction requires a three loop calculation.  
One dimension 2 operator, the Debye mass, appears here.
However, the large value of $m_D^2$ makes the influence of the $A_0$
field on the
thermodynamics of the gauge fields perturbative; to study their
thermodynamics, we will want to integrate out the $A_0$ field, and the
2 and 3 loop corrections to $m_D^2$ and 1 loop wave function
corrections to the $A_0$ field will only change the result of that
integration at order $\beta_L^{-3/2}$, as we discuss below.
Hence, we will not calculate these corrections here.
But we will need to know the Debye mass.  Since the bare value is zero,
$m_{D}^2$ equals the counterterm, which is computed at one loop in
\cite{FKRS}:
\begin{equation}
m_D^2 = \frac{2 N_c \Sigma}{\pi \beta_L} \;({\rm lattice}) \; = 
	\frac{N_c \Sigma \beta_L}{8 \pi} g^4 T^2 \; ({\rm physical}) \, .
\end{equation}

\begin{figure}
\centerline{\psfig{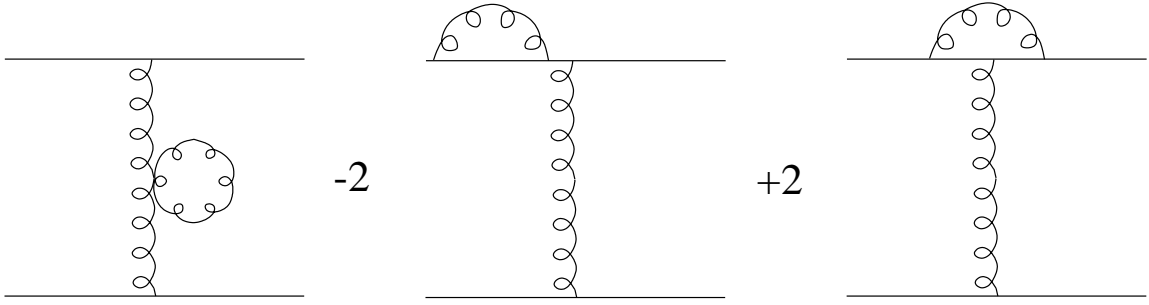}}
\caption{\label{renpic}
Examples of diagrams contributing to the renormalization of the gauge field.
The full effect of a gauge line propagating between scalar lines must 
match between theories; besides self-energy corrections, there are also
corrections at the vertex, and corrections because the vertex is
renormalized at the same time the scalar propagator is.}
\end{figure}

It only remains to find the correction to the plaquette
term, which must be multiplicative, ie the term
$\sum 1 - (1/2) {\rm Tr} U$ above is multiplied by
a wave function correction $Z_A$.  This can be absorbed into (or
understood as) a shift in
$\beta_L$.  Contributions to $Z_A$ arise both from self-energy and vertex
corrections, and are actually easiest to compute
in the theory with $N_s$ fundamental scalars added.
The gauge field is properly normalized if the full effect of a gauge
line propagating between scalar lines is the same in the lattice and
continuum theories.  The scalar wave function receives a renormalization
which, by minimal coupling, changes the strength of the scalar-gauge
vertex.  There are also loop corrections to the vertex and to the
propagator, illustrated in Figure \ref{renpic}.
What enters the calculation is the difference 
between the loop corrections to the gauge-scalar vertex and the scalar 
wave function, and the pure gauge and $A_0$
contributions to the gauge self-energy.  The
only differences between the SU(2) and SU($N_c$) calculations are the
group factors; no new diagrams or new momentum integrals appear.
Examining \cite{Oapaper}, one finds that the fundamental scalar corrections 
to the self-energy depend on $N_s {\rm Tr} T_a T_b = (N_s/2) \delta_{ab}$,
and that all
but one of the other diagrams have group factors proportional to
$N_c$ (if one takes the appropriate differences
between contributions to the scalar-vector vertex and the scalar wave
function correction).  The diagrams needed and their values are tabulated
there.

The one exception is a contribution to the gauge field
tadpole diagram arising from a term in the gauge field 4-point interaction,
which comes entirely from anticommutators of Lie algebra generators,
and has a group factor of
\begin{equation}
\frac{1}{3} \left( \frac{2}{N} \left[ \delta_{ab} \delta_{cd} + 
	\delta_{ac} \delta_{bd} + \delta_{ad} \delta_{bc} \right]
	+ d_{abe}d_{cde} + d_{ace}d_{bde} + d_{ade}d_{bce} \right)
\end{equation}
and a Lorentz dependence of $(1/2) \sum_{ij} F_{ij}^4$.  
This term is purely
nonrenormalizable operator and has no analog in the continuum theory; it
also dominates the correction between lattice and continuum theories.
To find the group factor for the contribution to $Z_A$
one contracts against $\delta_{cd}$; the $\delta_{ab}$ type
terms give $(2/3)(N + 1/N)$ and the $d_{abe}$ type terms give 
$(2/3)(N-4/N)$.  The space integral gives $1/(3 \beta_L)$.  

All told, the
contributions to the gauge field renormalization are 
($\xi = 0.152859$, $\Sigma = 3.17591$)
\begin{equation}
\beta_L (Z_A - 1) = 
\frac{2}{9}\left( 2N_c - \frac{3}{N_c} \right) 
	+ N_c \left( \frac{37 \xi}{12 \pi} - \frac{1}{9} \right) 
	+ N_c \left( \frac{\Sigma}{24 \pi} - \frac{\xi} {6 \pi} \right) 
	+ N_s \left( \frac{\Sigma}{48 \pi} - \frac{\xi}
	{12 \pi} \right) \, ,
\end{equation}
where the first term is from the contribution discussed above, the
second term is from all other gauge field contributions, the third
is from the (adjoint scalar) $A_0$ field contribution to the self-energy,
and the last term is if there are fundamental scalars present.
If one puts in a ``naive'' bare value of $\beta_{L,bare}$ then the 
simulation is equivalent to one with the appropriate $Z_A$ and
$\beta_{L,imp} = Z_A^{-1} \beta_{L,bare}$, which numerically is
\begin{equation}
\beta_{L,imp} = \beta_{L,bare} 
	- \left( \frac{4N_c}{9} - \frac{2}{3N_c} \right)
	- .0389 N_c - .0340 N_c - .0170 N_s
\label{betaimp} 
\end{equation}
where again the contributions are from the anticommuting part of the
tadpole diagram, other vector contributions, the $A_0$ field, and any
scalars present.  For $N_c=3$ and no fundamental scalars, we find
$\beta_{L,imp} = \beta_{L,bare} - 1.330$.  This gives us the correction
we need to convert from lattice to physical length scales.

Two comments are in order here.  First, the correction is totally
dominated by the tadpole term.  This behavior is typical in lattice
gauge theory.  It is caused by the compact nature of the connections
and it is the reason a perturbative matching between the lattice and
continuum theories is so necessary.  Second, 
the correction is larger for SU(3) than for SU(2).  Since the
diffusion constant depends on the fourth power of the conversion between
length scales, one must ensure that the matching is quite good.  If for
instance we assume that there are unknown $O(a^2)$ tadpole type
corrections of magnitude equal the square of the $O(a)$ correction, 
and we ask that these
corrections be at the $1\%$ level (which will still give $4 \%$
systematic errors), we need $\beta_L > 13$; we must work on quite
fine lattices.  Though this increases the numerical
demands, it is not all bad, since it means that the 
elementary plaquettes will be quite close to the identity and it should
be possible to make the connections quite smooth.  We will need this
property in the next subsection.

Now we have related the lattice theory to the continuum theory with
$A_0$ field, and with the wrong Debye mass.  We need to understand how
the value of $m_D^2$ modifies the gauge field thermodynamics, and this
is easiest done by integrating out the $A_0$ field.  In terms of the
natural 3-D length scale $1/g^2 T$, the Debye mass is $m_D^2 \sim
\beta_L (g^2 T)^2$, so it is indeed heavy enough to integrate out.  At
one loop, the modification to the gauge coupling is \cite{KLRS}
\begin{equation}
\bar{g}^2 = g^2 \left( 1 - \frac{N_c g^2 T}{48 \pi m_D} \right).
\label{gbar}
\end{equation}
This correction is formally $O(\beta_L^{-1/2})$.  Now recall that
$m_D^2$ receives two loop corrections which are $O(\beta_L^{-2})$ in
lattice units, or $O(g^4 T^2)$ in physical units.  One can see by
plugging $m_D^2 = ( A \beta_L + B ) g^4 T^2$ into Eq. (\ref{gbar}) that
this correction leads to an
$O(\beta_L^{-3/2})$ correction to $\bar{g}^2$.  Similarly, the
$O(\beta_L^{-1})$ correction to the $A_0$ wave function
renormalization can be absorbed by a rescaling of $A_0$ into a shift
in $m_D^2$ of order $O(m_D^2 \beta_L^{-1})$, also leading to an
$O(\beta_L^{-3/2})$ correction.  We will neglect these
$O(\beta_L^{-3/2})$ corrections.

In addition to the $O(\beta_L^{-1/2})$ correction we have just
mentioned, the integration over the $A_0$ field gives 
an $O(\beta_L^{-1})$ two loop correction, and
induces nonrenormalizable operators which affect physics at the
nonperturbative length scale by $O(\beta_L^{-3/2})$.  While the
nonrenormalizable operators are ignorable, The two loop
correction is parametrically as important as the
$O(\beta_L^{-1})$ correction to the lattice-continuum match which we
have just calculated.  However, we believe that it is numerically
much smaller.  This is because of the ``tadpole'' character of the
dominant 1-loop effects we have studied.  It can be argued that the
expansion parameter for the integration over the $A_0$ field is $g^2
T/ 4 \pi m_D \sim 0.1 \beta_L^{-1/2}$, whereas the ``tadpole''
corrections were $\sim 1 \times \beta_L^{-1}$.  Numerically, at 
$\beta_L = 16$, the one loop, $O(\beta_L^{-1/2})$ correction 
to $g^2$ from integrating out the $A_0$ field, Eq. (\ref{gbar}), 
is less than $1 \%$, while the $O(\beta_L^{-1})$ correction in
the lattice-continuum match is $\sim 8 \%$.  Hence, although the two loop
correction from integrating out the $A_0$ field is formally of the
same parametric order as corrections we include, it should be
numerically unimportant.  
Note that the numerical unimportance of this two loop effect is also
important
in the study of thermodynamic properties of SU(2) YMH theory using the
dimensional reduction program if one integrates out the $A_0$ field in
that program.  We will only make corrections due to
Eq. (\ref{betaimp}) and Eq. (\ref{gbar}) in this paper.


Finally, there are also $O(a)$ lattice corrections to 
Eq. (\ref{Edef}), which can be understood as corrections in the 
conversion between lattice and continuum time units.  These have not
been calculated, but it was argued in \cite{slavepaper} that they are
dominated by tadpole effects which are the same as those occurring in
the 
lattice-continuum match for the spatial gauge fields.  We will use the
prescription proposed there, leading to another $O(a)$ but probably
modest error.

The thermodynamics are now under control, and while there will still
be uncorrected $O(a)$ errors in $\kappa_1$, the large ``tadpole'' type
corrections are taken care of.

\subsection{winding number}

Now we discuss the extension of the winding number tracking technique 
of \cite{slavepaper} to SU($N_c$).

The idea of that paper is to keep track of a notional group valued 
scalar field $S$, with Hamiltonian 
\begin{equation}
H_S = \sum_{x,i}  \frac{N_c}{2} - \frac{1}{2} {\rm Re \, Tr} 
	S^{\dagger}(x) U_i(x) S(x+i) \, .
\end{equation}
One then evolves $S$ dissipatively and agressively to minimize $H_S$.  The
Chern-Simons number of a configuration with $S = I$ everywhere is 
approximated to be zero, and the total winding number change during
an evolution is tracked by gauge transforming to the gauge $S = I$ whenever
that gauge is {\it everywhere} smooth, ie there is no neighborhood 
where the connection matricies $U$ are far from the identity.  When the
winding number of the underlying gauge field 
configuration changes, then $S$ will
go through a period where it is not smooth somewhere, as it adjusts to
describe the new winding number state.  When it has returned to being
everywhere smooth, the gauge transform to $S=I$ is a large (but smooth)
transformation; we find its winding number and use it to increment the
cumulative winding number change to date.  The winding number 
of a gauge transformation is determined
with an algorithm which is essentially that of Woit \cite{Woit}.

Two things become more complicated when one goes to SU($N_c$).  The first
is the implementation of the dissipative algorithm for $S$.  The basic
element of the dissipative algorithm is to minimize $H_S$ with respect
to $S$ at one site $x$ \cite{Mandula}.  The easiest
way in SU(2) to find $S(x)$ which minimizes $H_S$ is to sum the parallel
transports of nearest neighbors, which will be a constant times
the desired element of SU(2), 
and to project the modulus to SU(2).  For SU($N_c$) the
sum of several group elements is not generally a multiple of a group
element, and one must orthogonally project to SU($N_c$) by a more complicated
algorithm.  First, scale the matrix so the modulus of its determinant
will be close to 1.  Call the resulting matrix $M$.  To project towards
U($N_c$), one repeatedly replaces $M \rightarrow (3/2) M - 
(1/2) M M^{\dagger} M$; if this process converges it gives the closest
element of U($N_c$) to the original matrix $M$.  Then one makes a U(1)
rotation by the angle $(-1/3) \arg {\rm Det} M$ to get to SU($N_c$).  If the
projection to U($N_c$) fails to converge, for instance because the 
slave field is varying wildly around the point, then one does not update
at this point; but in practice this essentially never happens.  The
algorithm to combine these elementary projections into an efficient 
quench is the same as in the SU(2) case \cite{slavepaper}.

The second complication is determining the winding number of a slave
field configuration.  For SU(2) there is a simple algorithm because
the group has the same dimension as the space.  Interpolating
$S(x)$ between lattice sites by a geodesic rule, $S$ becomes a map from 
$T^3$ to SU(2)$\cong S^3$, which are of the same dimension, and the 
winding number is just the oriented sum of times some fixed point in SU(2) is
covered \cite{Woit}.  There is an extension of this idea to SU(3) by
choosing a cannonical map of SU(3) with a dimension 4 subspace cut out 
into $S^3 \times S^5$ such that the $S^3$ part carries the relevant 
topological information \cite{Parisi}.  The 4 dimensional excision will
not generally be important since SU(3) is 8 dimensional and we are mapping
from a 3 dimensional space.

\begin{figure}[t]
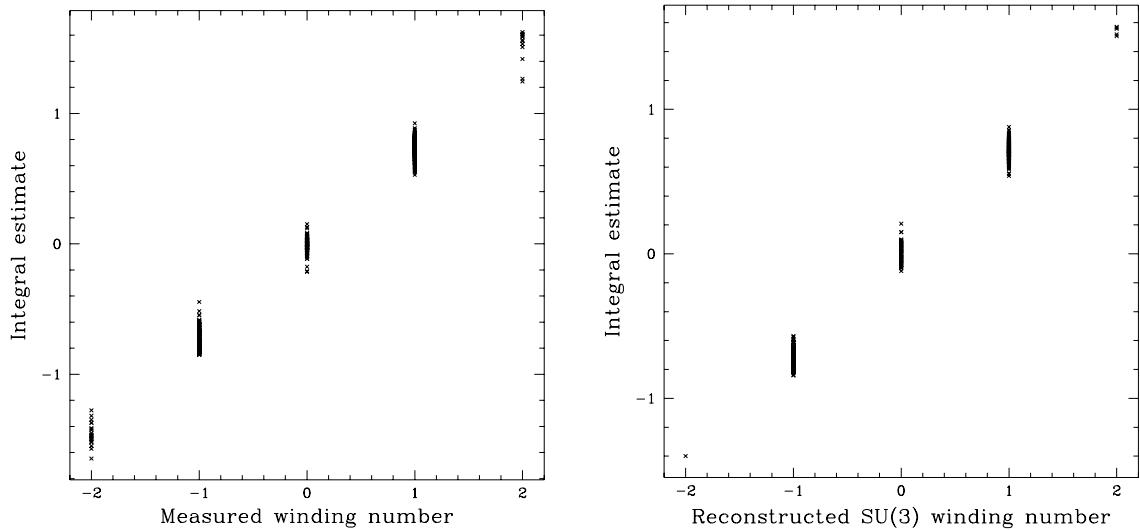

\centerline{
\mbox{\psfig{file=compSU2.epsi,width=2.8in}} \hspace{0.2in}
\mbox{\psfig{file=compSU3.epsi,width=2.8in}} }
\caption{\label{itworks}
The value of the integral, Eq. (\protect{\ref{NCSintegral}}), versus the
actual winding number for gauge transformations in an
SU(2) evolution at $\beta_L = 8$ on
a $20^3$ lattice, at left, shows that the integral 
can be used to unambiguously 
reconstruct the original winding number.  At right,
Eq. (\protect{\ref{NCSintegral}}) for gauge transformations in an SU(3) 
evolution on a $24^3$ grid at $\beta_L = 16$, plotted against the 
reconstructed winding number.  The values of the integral are clustered
with large breaks which makes the reconstruction unambiguous, even though
we have no direct integer measure of the winding number.}
\end{figure}

However, because the underlying fields are
quite smooth, it turns out there is an easier way to determine the
winding number.  One constructs the vacuum field obtained by gauge
transforming the naive $U=I$ vacuum by $S$, and then estimates its
$N_{CS}$ directly by integrating $\epsilon_{ijk} f_{abc} A^i_a A^j_b A^k_c$.
Defining 
\begin{equation}
A_i^a(x+i/2) = \frac{1}{2} {\rm Re Tr} -i \lambda^a U_i(x) 
	= \frac{1}{2} {\rm Re Tr} S^{\dagger}(x+i) (-i \lambda^a) S(x)
\end{equation}
and 
\begin{equation}
A_i^a(x) = \frac{A_i^a(x+i/2) + A_i^a(x-i/2) }{2} \, ,
\end{equation}
the integral is
\begin{equation}
N_{CS} = \frac{1}{2 \pi^2} \sum_x f_{abc} A^a_1(x) A^b_2(x) A^c_3(x) \, .
\label{NCSintegral}
\end{equation}
The result is not gauge invariant and will not be an integer.  However,
if the slave field is smooth, then the result will be close to an integer;
if the slave field used in a gauge transformation is always suitably smooth
then it will be possible to unambiguously reconstruct what integer
the above integral was ``trying to give us''.
We have tested this idea with SU(2), where it is possible to compare
the value of the integral to the (integer) winding number determined 
topologically.  This is illustrated in Figure \ref{itworks}; as seen there,
the values from Eq. (\ref{NCSintegral}) always understate the winding 
number, but by a fairly consistent amount.  The values obtained for
gauge transformations of one winding number do not overlap those which
arise from another winding number, so one can unambiguously reconstruct
the winding number of gauge transformations from the values of Eq. 
(\ref{NCSintegral}).
To implement the same
idea in SU(3), one just writes the time and value of Eq. (\ref{NCSintegral})
every time a gauge transformation is made.  Plotting the values of the
integral which occurred, one finds the appropriate breaks and can then
reconstruct the winding number changes, also illustrated in 
Figure \ref{itworks}.
There is never any difficulty in the reconstruction for
the lattice spacing and volumes used in this paper, though presumably this
technique should start to have problems on very coarse lattices or
large volumes.

\section{Numerical results}

It appears from theoretical arguments \cite{ArnoldYaffe,Arnoldlatt} and 
numerical results \cite{slavepaper} that the $N_{CS}$ diffusion constant
in SU(2), $\kappa_{ws}$, depends on the lattice spacing, because
finer lattices have more hard modes contributing to hard thermal
loops \cite{Smilga}.  Since it is more numerically expensive to study
SU(3) it makes the most sense to try to come to grips with this
problem, and with the problem of trying to include hard thermal loops
properly, in the SU(2) theory.  Hence we will determine the 
lattice value of $\Gamma_{ss}$ at only one lattice spacing, albeit a
fairly fine one.  It should be possible to use this value to establish
the ratio $\Gamma_{ss}/\Gamma_{ws}$, up to errors from how $\Gamma$
approaches the large $\omega_{pl}$ scaling regime.
The absolute value of $\Gamma_{ss}$
can also be estimated using the arguments in \cite{Arnoldlatt}, but
the systematic errors will be bigger here.

It is not known what finite volume systematics
may occur in the calculation of $\Gamma_{ss}$, so we measure it
on a range of (cubic toroidal) lattices.  Our results are for 
$\beta_{L,bare} = 16$ and for $8^3,$ $12^3$, $16^3$, $24^3$, and
$32^3$ lattices.  The results are tabulated in Table \ref{onlytable},
which presents $\kappa_{ss}$, where $\kappa_{ss}$ is defined in
Eq. (\ref{naivescaling}).  Naively, one would convert from lattice to 
continuum units by 
\begin{equation}
\kappa_{ss} = {\rm diffusion \; constant} ({\rm lattice \; units}) 
	\times (\pi \beta_{L,bare})^4 \, ,
\end{equation}
but we have used the thermodynamic corrections derived in the last
section, as described in \cite{slavepaper}.

The largest 3 lattices are statistically compatible, so we have
achieved the large volume limit at least at the level of statistics
obtained here.  The winding number changes on the smallest lattice
were all immediately followed by a winding number change of opposite
sign; this might be the system getting almost up to a half integer
$N_{CS}$ state and then turning back, and does not represent
a permanent change to the underlying vacuum winding number.  We
have only an upper limit for the diffusion constant at that lattice
spacing.

\begin{table}[t]
\centerline{ \mbox{
\begin{tabular}{|c|c|c|} \hline
 \hspace{0.2in} $\beta_{L,bare}$ \hspace{0.2in}  &
 \hspace{0.2in}   $N$  \hspace{0.2in}  & 
 \hspace{0.3in}  $\kappa_{ss}$  \hspace{0.3in}  \\ \hline
 16 & 8  &  $.00 \pm .07$ \\ \hline
 16 & 12 &  $4.0 \pm 0.8$ \\ \hline
 16 & 16 &  $8.3 \pm 1.1$ \\ \hline
 16 & 24 &  $8.8 \pm 1.4$ \\ \hline
 16 & 32 &  $8.0 \pm 1.2$ \\ \hline
\end{tabular} } }
\caption{\label{onlytable}
Dependence of $\kappa_{ss}$ on lattice volume, showing the approach to
an infinite volume limit.}
\end{table}

We would like to use this lattice rate to estimate the rate in the
physical quantum theory.  As we have argued, thermodynamic errors, ie
errors in $\kappa_1$, are under control and should be smaller than our
statistical errors.  However, the same cannot be said of systematics
in the dynamics, ie in $\kappa_2$.  The problem of relating the
classical lattice theory and the real, continuum quantum one has
recently been studied by Arnold \cite{Arnoldlatt}.  He argues that, in
the large HTL effect regime, the evolution of infrared magnetic fields
is overdamped on time scales longer than $1/g^2 T$, and the strength
of the damping sets $\kappa_2$.  The damping occurs because $\vec{E}$
fields set up currents of ``hard'' modes; the currents propagate; and
they enter the soft mode equations of motion somewhere else.  Hence,
the infrared fields at one location ``feel'' the electric fields along
the past light cone, due to interactions with hard modes.  This is the
physics of the hard thermal loops.  However, the distribution of hard
modes on the lattice is very anisotropic; so will be the conveying of
information by the hard modes; and so will be the damping of the
infrared fields.  Arnold argues that one can make an approximate match
between classical lattice and continuum quantum values of $\kappa_2$
by taking an angular average of the damping strength.  There are
systematics associated with this, which he estimates conservatively as
being on order $30 \%$.  There is another systematic
because the lattice system is probably not deep in the strong damping
regime; neither is the continuum quantum theory at realistic
$\alpha_s$. 

Using Arnold's proposed match of damping coefficients, we find that
$\Gamma_{ss}$ for SU(3) with six flavors of quarks is (including the
factor of $1/2$ to go from the diffusion constant to the response
coefficient) 
\begin{equation}
\label{Arnoldway}
\Gamma_{ss} = ( 108 \pm 15_{\rm stat} \times 2^{\pm 1 \; {\rm
	syst }} ) \alpha_s^5 T^4 \qquad
	( m_D^2 = 2 g_s^2 T^2 ) \, .
\end{equation}
Here I have made a conservative estimate of the systematic errors to
be about a factor of 2.

The ratio $\Gamma_{ss} / \Gamma_{ws}$ in the formal small $\alpha$
limit has smaller systematics because the problem from the anisotropy
of the damping coefficient is common to SU(2) and SU(3).  Corrections
to the large HTL limit are probably of different magnitude for the two
theories, though they are presumably of the same sign.  We can
estimate them by fitting the SU(2) data at different values of
$\beta_L$ to the functional form $A \beta_L^{-1} ({\rm leading}) + B
\beta_L^{-2} ( {\rm correction} )$ and seeing how large the
extrapolation from $\beta_L = 16$ to $\beta_L = \infty$ is.  From the
data in \cite{slavepaper}, which has the same thermodynamic
improvements as here and a topological definition of $N_{CS}$, we find
that the extrapolation is a $21 \%$ correction.  (The fit is
startlingly good:  $\chi^2 =.42$ for 5 points and 2 fitting
parameters.)  If the (unknown)
difference between the SU(2) and SU(3) extrapolations is on order the
same size as the SU(2) extrapolation, then taking the ratio of
$\beta_L = 16$ data and using the size of the extrapolation to give
the systematic error gives
\begin{equation}
\frac{\Gamma_{ss}}{\Gamma_{ws}} = \left( 10.7 \pm 1.5_{\rm stat} 
	\pm 2.3_{\rm syst}
	\right) \left( \frac{\alpha_s}{\alpha_w} \right)^5
	\frac{m_{Dw}^2 g_s^2}{m_{Ds}^2 g_w^2} \, .
\label{gammaratio}
\end{equation}
(The direct ratio of $\beta_L = 16$ data is $7.1$, but remember that
$m_{DL}^2 \propto N_c$.)

To evaluate this we need $\alpha_s$ and $\alpha_w$ in the
dimensionally reduced 3-D theory.
Using $T_c = 100$GeV and
$\alpha_s(M_Z,\overline{\rm MS}) = 0.118$, we ran $\alpha_s$ to
the renormalization point $\mu = 7.06 T$ using the two loop
renormalization group equation, and used this value as input in
Eq. (146) of \cite{KLRS}, adapted to six flavor SU(3), 
to find $\alpha_{s,DR}$.  This procedure should
minimize two loop errors.  The result is $\alpha_{s,DR} = 0.086$.  We
took $\alpha_w$ from Figure 7 of \cite{KLRS}, at $T_c = 100$GeV and
$m_H = 70$GeV:  $\alpha_{w,DR} = 1/31.7$.  Both values and particularly
the value for $\alpha_{s,DR}$ are smaller than we are used to; this is
because the dimensional reduction procedure sets the coupling roughly
to the value at $\mu(\overline{\rm MS}) \sim 7 T$.  
Using these values, $m_{Dw}^2
= 11 g_w^2 T^2/6$, and $m_{Ds}^2 = 2 g_s^2 T^2$, evaluating 
Eq. (\ref{gammaratio}) gives
$\Gamma_{ss} / \Gamma_{ws} \sim 1500$.\footnote{Technically we should 
not just use $\alpha_{s,DR}$ but should also include the correction,
Eq. (\protect{\ref{gbar}}).  But even after taking the fifth power of 
$\alpha_s$, this correction is less than $10 \%$.  This is another
example of how well behaved the integration over the $A_0$ field
turns out to be.}  This is much smaller
than the value we would get using $\alpha_s = \alpha_s(M_Z) \simeq
0.118$.  This large renormalization point dependence makes one nervous
that subleading contributions in $\alpha_s$ may be non-negligible.

\section{Conclusion}

We have calculated the diffusion constant for SU(3) Chern-Simons number at
weak coupling (ie, high temperatures) by classical, lattice methods, using
a topological definition of $N_{CS}$, and find that, at equal values of
coupling and hard thermal loops, the diffusion constant is larger than
in SU(2) by an order of magnitude.  Since in the standard model above
the electroweak phase transition, the strong coupling constant is quite
a bit larger than the weak coupling constant, the actual ratio of
winding number diffusion rates for SU(3) and SU(2) is very large.

Since the hard thermal loop effects induced by the hard lattice modes
are different from those which would occur for ultrarelativistic
particles \cite{Smilga},
we do not know with certainty how to convert the diffusion constant
of the classical lattice system into the diffusion constant for the 
physical quantum system, and cannot establish the time constant $\tau$ 
with which a chiral quark number is damped by strong sphaleron processes
in the plasma.  But using the estimate, Eq. (\ref{Arnoldway}), we get
$\tau \sim 80/T$.  This is slow enough to allow quarks to escape the
bubble wall before chiral quark number is destroyed, but it is still
over 3 orders of magnitude faster than the rate at which chiral 
quark number is converted
into baryons through weak sphaleron processes.

We should mention an interesting case where strong sphalerons are
less important, which is
for theories with a stop squark light enough to develop
a condensate just before the electroweak phase transition.  The 
thermodynamics of this model have been considered recently \cite{Lainestop}
and it apparently provides an especially strong phase transition.  Since
color is broken from SU(3) to SU(2) in the symmetric phase, strong
sphalerons only proceed at the SU(2) rate for two of the colors and 
will only erase chiral quark number in the third color at an exponentially
small rate.  Of course, strong processes will mix chiral quark number
between the two unbroken and one broken color, but the suppression of
the rate by a factor of 10 is significant when the ratio $\Gamma_{ss} /
\Gamma_{ws}$ is important to the final baryon number abundance.

\centerline{Acknowledgements}

I would like to thank Alex Krasnitz, James Hetrick, and Peter Arnold
for useful conversations or correspondence.  
This work was supported under NSF contract NSF -- PHY96-00258.


\begin{thebibliography}{99}
\bibitem{Sakharov} A. Sakharov, JETP Lett. {\bf 6} (1967) 24.
\bibitem{various} L. McLerran, M. Shaposhnikov, N. Turok, and M.
	Voloshin, Phys. Lett. {\bf B 256} (1991) 451; A. Cohen, D. Kaplan,
	and A. Nelson, Phys. Lett. {\bf B 263} (1991) 86;
	M. Dine and S. Thomas, Phys. Lett. {\bf B 328} (1994) 73.
\bibitem{two} A. Nelson, D. Kaplan, and A. Cohen,
	Nucl. Phys. {\bf B 373} (1992) 453; 
	M. Joyce, T. Prokopec, and
	N. Turok, Phys. Rev. {\bf D 53} (1996) 2958.
\bibitem{CKNdiffusion} A. Cohen, D. Kaplan, and A. Nelson, Phys.
	Lett. {\bf B 336} (1994) 41.
\bibitem{SSearly} L. McLerran, E. Mottola, and M. Shaposhnikov,
	Phys. Rev. {\bf D 43} (1991) 2027.
\bibitem{Giudice} G. Giudice and M. Shaposhnikov, Phys. Lett.
	{\bf B 326} (1994) 118.
\bibitem{KhlebShap} S. Khlebnikov and M. Shaposhnikov, Nucl. Phys.
        {\bf B308} (1988) 885.
\bibitem{RubakShap2}  V. Rubakov and M. Shaposhnikov,
	Phys. Usp. 39 (1996) 461-502, 
	(Usp. Fiz. Nauk 166 (1996) 493-537.).
\bibitem{Moore1} G. D. Moore, Nucl. Phys. {\bf B 480} (1996) 657.
\bibitem{GrigRub} D. Grigorev and V. Rubakov, Nucl. Phys. {\bf B 299}
	(1988) 248.
\bibitem{Ambjornetal} J. Ambj{\o}rn, T. Askgaard, H. Porter, and M.
             Shaposhnikov, Nucl. Phys. {\bf B 353} (1991) 346.
\bibitem{AmbKras} J. Ambj{\o}rn and A. Krasnitz, Phys. Lett. {\bf B 362}
	(1995) 97.
\bibitem{Oapaper} G. D. Moore, Nucl. Phys. {\bf B 493} (1997) 439.
\bibitem{ArnoldYaffe} P. Arnold, D. Son, and L. Yaffe, Phys. Rev.
	{\bf D 55} (1997) 6264;
	P. Huet and D. Son, Phys. Lett. {\bf B393} (1997) 94.
\bibitem{Arnoldlatt} P. Arnold, Phys. Rev. {\bf D 55} (1997) 7781.
\bibitem{slavepaper} G. D. Moore and N. Turok, PUPT-1681, hep-ph/9703266.
\bibitem{Smilga} D. Bodeker, L. McLerran, and A. Smilga, Phys.
	Rev. {\bf D 52} (1995) 4675.
\bibitem{Hetrick} James Hetrick, private communication.
\bibitem{AmbKras2} J. Ambj{\o}rn and A. Krasnitz, hep-ph/9705380.
\bibitem{Woit} P. Woit, Phys. Rev. Lett. {\bf 51} (1983) 638;
	Nucl. Phys. {\bf B 262} (1985) 284.
\bibitem{KLRS} K. Kajantie, M. Laine, K. Rummukainen, and M.
	Shaposhnikov, Nucl. Phys. {\bf B 458} (1996) 90.
\bibitem{TangSmit} W. Tang and J. Smit, Nucl. Phys. {\bf B 482} (1996) 265.
\bibitem{KogutSusskind} J. Kogut and L. Susskind, 
	Phys. Rev. {\bf D 11} (1975) 395.
\bibitem{FKRS} K. Farakos, K. Kajantie, K. Rummukainen, 
	and M. Shaposhnikov, Nucl. Phys. {\bf B 442}
	(1995) 317.
\bibitem{Mandula} J. Mandula and M. Ogilvie, Phys. Lett {\bf B 185} (1987)
	127.
\bibitem{Parisi} G. Parisi and F. Rapuano, 
	Phys. Lett. {\bf B 152} (1985) 218.
\bibitem{Lainestop} D. B{\"o}deker, P. John, M. Laine, and M. Schmidt,
	HD-THEP-96-56, hep-ph/9612364. 
\end{thebibliography}
\end{document}